\documentclass[conference,a4paper]{IEEEtran}
\usepackage[utf8]{inputenc} 
\usepackage[T1]{fontenc}
\usepackage{url}              
\usepackage{cite}             

\usepackage[cmex10]{amsmath}  
\usepackage{mleftright}       
\mleftright                   

\usepackage{graphicx}         
\usepackage{booktabs}         

 \usepackage[caption=false,font=footnotesize]{subfig}

\usepackage{amsfonts, bm}
\usepackage{amsmath}
\usepackage{amssymb}
\usepackage{graphicx}
\usepackage{prettyref}
\usepackage{multirow, tikz, pgf, pgfplots}
\usepackage{caption, float}
\usepackage{xcolor}
\usepackage[utf8]{inputenc}
\usetikzlibrary{shapes.geometric}
\usetikzlibrary{arrows.meta,arrows}

\newcommand{\bbR}{\mathbb R}
\newcommand{\bbC}{\mathbb C}

\newcommand{\bbE}{\mathbb E}
\newcommand{\bbP}{\mathbb P}
\def\EE{\bbE}

\newcommand{\TV}{\mathsf{TV}}
\newcommand{\indc}[1]{{\mathbf{1}_{\left\{{#1}\right\}}}}

\newrefformat{lmm}{Lemma~\ref{#1}}
\newrefformat{fig}{Fig.~\ref{#1}}
\newrefformat{app}{Appendix~\ref{#1}}
\newrefformat{cor}{Corollary~\ref{#1}}

\newtheorem{theorem}{Theorem}
\newtheorem{corollary}{Corollary}[theorem]
\newtheorem{lemma}[theorem]{Lemma}

\newtheorem{remark}{Remark}

\newcommand{\stepa}[1]{\overset{\rm (a)}{#1}}
\newcommand{\stepb}[1]{\overset{\rm (b)}{#1}}
\newcommand{\stepc}[1]{\overset{\rm (c)}{#1}}
\newcommand{\stepd}[1]{\overset{\rm (d)}{#1}}

\newcommand{\calN}{{\mathcal{N}}}

\newcommand{\calX}{{\mathcal{X}}}

\newif\ifmapx
{\catcode`/=0 \catcode`\\=12/gdef/mkillslash\#1{#1}}
\edef\jobnametmp{\expandafter\string\csname main_apx\endcsname}
\edef\jobnameapx{\expandafter\mkillslash\jobnametmp}
\edef\jobnameexpand{\jobname}
\ifx\jobnameexpand\jobnameapx
\mapxtrue
\else
\mapxfalse
\fi

\ifmapx
\title{Comparing Poisson and Gaussian channels (extended)}
\else 
\title{Comparing Poisson and Gaussian channels}
\fi

\author{%
	   \IEEEauthorblockN{Anzo Teh and Yury Polyanskiy}
	   \IEEEauthorblockA{EECS, MIT\\
		                     Cambridge, Massachusetts\\
		                     USA\\
		                     \{anzoteh, yp\}@mit.edu}
	 }

\def\Gsn{\mathsf{Gsn}}
\def\Poi{\mathsf{Poi}}

\begin{document}
    \maketitle
    
\begin{abstract} 
	Consider a pair of input distributions which after passing through a Poisson
channel become $\epsilon$-close in total variation. We show that they must necessarily then be
$\epsilon^{0.5+o(1)}$-close after passing through a Gaussian channel as
well. In the opposite direction, we show that distributions inducing $\epsilon$-close outputs over the Gaussian channel
must induce $\epsilon^{1+o(1)}$-close outputs over the Poisson. This quantifies a well-known
intuition that ``smoothing''
induced by Poissonization and Gaussian convolution are similar. As an application, we improve a
recent upper bound of Han-Miao-Shen'2021 for estimating mixing distribution of a Poisson mixture in Gaussian optimal
transport distance from $n^{-0.1 + o(1)}$ to $n^{-0.25 + o(1)}$.
\end{abstract}

    \section{Introduction}

    Fix three positive parameters $a,\sigma, \gamma > 0$ and consider \textit{two channels} with a common
    input space $\calX=[0,a]$. The first channel, denoted $\Gsn_\sigma$, acts on input $X=x_0$ by outputting
    	$Y_G \sim \calN(x_0,\sigma^2)$. 
    The second channel, denoted $\Poi_{\gamma}$, acts by outputting
    	$Y_P \sim \text(Poi)(\gamma x_0)$.
    Note that the output spaces of these two channels are very different. For the first one $Y_G
    \in \mathbb{R}$ and for the second one $Y_P \in \mathbb{Z}_+$. When $X \sim \pi$ we denote by
    $\Gsn_\sigma \circ \pi$ and $\Poi_{\gamma} \circ \pi$ the laws of $Y_G$ and $Y_P$, respectively. Despite the
    fact that these probability measures live on different spaces, we can view either of
    them as a kind of ``smoothed'' version of $\pi$, which destroys small local variations in $\pi$. 
    One may wonder, thus, whether one can perturb a fixed $\pi$ in such a way that the
    perturbation, while invisible after passing through Poisson channel, is apparent after passing
    through the Gaussian one. 
    In this work, we answer this in the negative and provide quantitive bounds. Specificially, we show that whenever two measures $\pi_1$
    and $\pi_2$ have total variation distance $\epsilon$ after Poisson smoothing, they must necessarily also be close after
    Gaussian smoothing (within total variation almost $O(\sqrt{\epsilon})$), and an even better
    bound in the opposite direction. Informally speaking, this demonstrates that the information embedded in local
    variations of $X$ is destroyed similarly by both channels. 

    Besides independent interest, our results have various applications. One could be in the
    domain of \textit{covert communication}~\cite{bash2013limits}, where coded distribution is
    supposed to have low total variation distance from a pure noise (our result compares these
    tasks over two channels). However, our original motivation lies in the domain
    of \textit{Gaussian optimal transport} (GOT) introduced in~\cite{goldfeld_convergence_2020}.
    We recall that a $\sigma$-GOT distance is defined as 
    \begin{IEEEeqnarray}{Cl}\label{eq:got}
    	&W_1^{(\sigma)}(\nu,\mu) 
    	\nonumber\\
    	= &\inf_{P_{A,B}} \left\{ {\EE[|A-B|]}: A \sim \Gsn_\sigma\circ \nu, B \sim \Gsn_\sigma
    \circ \mu\right\}\,,
    \end{IEEEeqnarray}  
    with infimum over all possible joint distributions $P_{A,B}$ with given marginals.
    When $\sigma=0$ this corresponds to the standard Wasserstein distance and is denoted by $W_1$
    without the superscript. It is known that
    estimating a distribution (supported on $[0,1]^d$) in Wasserstein distance is rather slow
    (typically, at rate $n^{-1/d}$ from $n$ iid samples). If, however, one is interested in only
    recovering distribution up to features of scale $\sigma$, then estimation metric could arguably
    be replaced by $W_1^{(\sigma)}$. It turns out that estimating in the latter can be done at
    much faster rates. 

    One example of this phenomena, and a second motivation for this work, is a result
    of~\cite{han_nonparametric_2021}, who showed that estimating $\pi$ from $n$ iid samples of $\Poi_{\gamma}\circ
    \pi$ while essentially impossible~\cite{miao_fisher-pitman_2021}  in $W_1$ (rate being $\mathrm{polylog}(n)$) can be
    done in GOT at a polynomial rate of (almost) $n^{-0.1}$. Our channel comparison analysis
    paired with a recent bound of~\cite{jana_optimal_2022} improves the
    estimate to (almost) $n^{-1/4}$. 

    From the technical side, our innovation is bringing the complex-analytic tools, previously
    used for Poisson-type problems in~\cite{MS13,polyanskiy_sample_2017,dual2-2019,jana2020extrapolating}
    to bear on this channel comparison question.
    With this brief outline, we proceed to formal statements next.

    \textbf{Notation.} $\lesssim$ and $\gtrsim$ denote inequalities up to absolute constants (in
    particular, these constants do not depend on the problem parameters $a,\sigma$). Similarly,
    $O_{a,\sigma, \gamma}(1)$ and 
    $o_{a,\sigma, \gamma}(1)$ denote quantity that stays bounded or vanishes, but depends on $a,\sigma, \gamma$. $\log$ denotes a
    base-$e$ logarithm. 
    When doing summation or integral, we will denote $\pi(t)$ as the probability mass (or density) function of distribution $\pi$ at $t$. 

\section{Main results}

    Throughout the paper, we restrict ourselves to priors of bounded support. 
    That is,  we denote $\mathcal{P}([0,a])$ the set of all probability distributions supported on $[0, a]$. 
    In addition to $W_1^{(\sigma)}$, $W_1$ that were already defined, we also recall definition of
    $\TV$ and Hellinger for two distributions $P, Q$ as follows ~\cite[(7.3), (7.5)]{polyanskiy_information_2022}. 
    \begin{IEEEeqnarray}{C}
    	\TV(P, Q) \triangleq \bbE_{Q}\left[\left|\frac{dP}{dQ} - 1\right|\right] =\frac 12 \int |dP - dQ|,
    \end{IEEEeqnarray}
    \begin{IEEEeqnarray}{C}
    	H^2(P, Q) \triangleq \bbE_{Q}\left[\left(1 - \sqrt{\frac{dP}{dQ}}\right)^2\right] =\int(\sqrt{dP} - \sqrt{dQ})^2.
    \end{IEEEeqnarray}
    
\subsection{Comparison of Poisson and Gaussian channels}

    \begin{theorem}\label{thm:poi_gauss} There exists $c=c(a, \sigma, \gamma)>0$ such that for any
    	$\pi_1, \pi_2\in\mathcal{P}([0,a])$ we have 
    	\begin{IEEEeqnarray*}{Cl}
    		&\TV(\Poi_{\gamma}\circ \pi_1, \Poi_{\gamma}\circ \pi_2)\le \epsilon
    		\nonumber\\
    		\implies & \TV(\Gsn_{\sigma}\circ\pi_1, \Gsn_{\sigma}\circ\pi_2)\le 
    		c(a, \sigma, \gamma)\sqrt{\epsilon}t_{a, \sigma, \gamma}(\epsilon)
    	\end{IEEEeqnarray*}
        where $t_{a, \sigma}(\epsilon)=\epsilon^{o(1)}$ as $\epsilon \to 0$, and more explicitly, 
	we have
		$$ t_{a, \sigma, \gamma}(\epsilon) = 
        		\frac{\ell_\epsilon^{3/4} }{\sqrt{\log \ell_\epsilon} }
			e^{({\log a - \log(\sigma^2)  - \log(\gamma)\over 2} + o(1)) {\ell_\epsilon \over \log \ell_\epsilon}}
			\,, \; \ell_\epsilon = \log {1\over \epsilon} \,.$$
	
    \end{theorem}
    
    In the opposite direction, we have the following result.
    \begin{theorem}\label{thm:gauss_poi}
    	There exists $c=c(a, \sigma, \gamma)>0$ such that for any
    	$\pi_1, \pi_2\in\mathcal{P}([0,a])$ we have 
    	\begin{IEEEeqnarray*}{Cl}
    		&\TV(\Gsn_{\sigma}\circ\pi_1, \Gsn_{\sigma}\circ\pi_2)\le \epsilon 
    		\nonumber\\
    		 \implies &
    		\TV(\Poi_{\gamma}\circ\pi_1, \Poi_{\gamma}\circ\pi_2)\le c\epsilon
		e^{3\gamma\sigma\sqrt{2\log\frac{1}{\epsilon}}}.
    	\end{IEEEeqnarray*}
    \end{theorem}

\begin{remark}
    	We consider a simple example on how $\TV$s of Gaussian and Poisson mixtures behave.
    	Let $\pi_1=\delta_t$ and $\pi_2=\delta_{t+\epsilon}$ for some small $\epsilon > 0$, 
    	and $0 < t < a - \epsilon$, 
    	where $\delta$ is the dirac-delta distribution. 
    	Then 
    	$\TV(\mathcal{N}(t, 1), \mathcal{N}(t+\epsilon, 1))=\epsilon \cdot (\frac{1}{\sqrt{2\pi}} + o(1))$; 
    	while 
    	\[
    	\frac {\exp(-t)}{2}(1 - \exp(-\epsilon))\stepa{\le} \TV(\Poi(t), \Poi(t + \epsilon))\stepb{\le} \epsilon
    	\]
    	with (a) by comparing the PMF at 0 and (b) by \cite[(2.2)]{adell2006exact}. 
    	Since the TV of both channels are of $\Theta(\epsilon)$, the exponent of $\epsilon$ in \prettyref{thm:poi_gauss} and \prettyref{thm:gauss_poi} cannot exceed 1. 
    \end{remark}

\subsection{Application to Gaussian optimal transport }
    
    Next, we discuss statistical applications of the results above.
    Consider the problem of estimating the distribution $\pi$ supported on
    $[0,a]$ from $n$ iid indirect observations $Y_P \sim \text{Poi}(X)$, $X\sim \pi$. 
    Here we denote the shorthand notation $\Poi\triangleq \Poi_1$. One can pose
    different questions related to estimating $\pi$. For example, while estimating $\pi$
    in TV is impossible, it can be estimated, for example, in Wasserstein $W_1$ distance, albeit
    at a slow rate. Specifically, \cite{han_nonparametric_2021} and \cite{miao_fisher-pitman_2021} show that  
    \begin{equation}
        \inf_{\hat {\pi}}\sup_{\pi\in\mathcal{P}([0,a])}
        \bbE[W_1(\pi, \hat{\pi})]=\Theta_a\left(\frac{\log \log n}{\log n}\right)\,.
    \end{equation} 

    Despite the poor performance of estimation of mixing distribution $\pi$ in this nonparametric
    inverse problem, 
    estimation of the mixture distribution $\Poi\circ\pi$ can be done at an almost  parametric rate. 
    Several different estimators $\hat \pi$, including Non-Parametric Maximum Likelihood Estimator (NPMLE),
    minimum Hellinger distance and minimum $\chi^2$-distance, were shown by 
    \cite{jana_optimal_2022} to achieve an estimation rate (in Hellinger distance) given by
    \begin{equation}\label{eq:jana}
        \sup_{\pi \in\mathcal{P}([0,a])} \bbE[H^2(\Poi\circ\pi, \Poi\circ{\hat{\pi}})]
        \le O_a\left(\frac{\log n}{n\log \log n}\right)\,,
    \end{equation}
    %
    where the Hellinger squared distance $H^2$ was defined above. 
    It was shown previously in~\cite[Appendix E]{eb2021} that this estimation rate cannot be
    improved. 
    A previous result by \cite[Proposition 3.1]{lambert_asymptotic_1984} also shows convergence of $\chi(\Poi\circ\hat{\pi}\|\Poi\circ\pi)$ at the rate of faster than $n^{-(1/2 - \tau)}$ for any $\tau > 0$. 

    Finally, estimation under the GOT distance~\eqref{eq:got} was considered recently. Specifically, \cite[Theorem 3.1.]{han_nonparametric_2021} states that for any $0 < c < 0.1$, 
    there exists a constant $C= C(\sigma, a, c)$ such that the NPMLE solution $\hat{\pi}$
    satisfies 
    \begin{equation}
    	\sup_{\pi\in\mathcal{P}([0,a])}\bbE [W_1^{(\sigma)}(\pi, \hat{\pi})]\le Cn^{-c}\,.
    \end{equation}
    We note that~\cite{han_nonparametric_2021} presents results for non-Poisson channels as well,
    but for the Poisson channel $c=0.1-o(1)$ is the rate obtained therein, cf.~\cite[Remark 3.2]{han_nonparametric_2021}.
    
    Here we improve this result as follows.
    
    \begin{corollary}\label{cor:got}
    For any $0 < c < \frac{1}{4}$, 
    there exists a constant $C= C(\sigma, a, c)$ such that the NPMLE solution $\hat{\pi}$ of the Poisson mixture attains rate 
    \begin{equation}
    	\sup_{\pi\in\mathcal{P}([0,a])}\bbE [W_1^{(\sigma)}(\pi, \hat{\pi})]\le Cn^{-c}.
    \end{equation}
    Furthermore, if $a\le \sigma^2\gamma$, then the right-hand side can be replaced with $n^{-1/4}\mathsf{Polylog}(n)$.
    \end{corollary}
    
    \ifmapx
    We only sketch the main steps here; the complete proof is in \prettyref{app:proofs_appendix}. 
    \else
    We only sketch the main steps here. 
    \fi
    First, by the standard bounds, e.g.~\cite[(7.20)]{polyanskiy_information_2022} we have $\TV(P, Q)\le H(P, Q)$ and thus by Cauchy-Schwarz and~\eqref{eq:jana} we have 
   \begin{equation*}
    \sup_{\pi\in\mathcal{P}([0, a])}\bbE[\TV(\Poi\circ\pi, \Poi\circ\hat{\pi})]
    =O_a\left(\frac{1}{\sqrt{n}}\cdot\sqrt{\frac{\log n}{\log \log n}}\right) 
   \end{equation*}
   Next, we leverage \prettyref{thm:poi_gauss} to get:
   \begin{equation}
    \sup_{\pi\in\mathcal{P}([0, a])}\bbE[
    \TV(\Gsn_{\sigma}\circ\hat\pi, \Gsn_{\sigma}\circ\pi)]
    = O_{a, \sigma}\left(n^{-1/4 + o(1)} \right) \label{eq:xxd1}   
   \end{equation}
    which, in the case where $a\le \sigma^2\gamma$, 
    $n^{o(1)}$ is actually $\mathsf{Polylog}(n)$ given that 
    $t_{a, \sigma, \gamma}(\epsilon)$ in \prettyref{thm:poi_gauss} is $\mathsf{Polylog}(\frac{1}{\epsilon})$. 
    
    \ifmapx
    For the next step we need the following estimate, to be proven in \prettyref{app:proofs_appendix}. 
    \else 
    For the next step we need the following estimate. 
    \fi 
   \begin{lemma}\label{lmm:tv_w1}
    	There exists $c_1= c_1(a,\sigma)$ such that 
    	for all $\pi_1, \pi_2\in\mathcal{P}([0,a])$ and for all $\delta > 0$ we have\footnote{We remark that the bound is likely not tight, as for example when $\sigma = 0$, we can easily get a better bound of $W_1(\pi_1, \pi_2)\le \frac{a\delta}{2}$ \cite[Theorem 6.15]{optimal.transport.old.new}.}
    	\begin{equation}
    		\TV(\Gsn_{\sigma}\circ\pi_1, \Gsn_{\sigma}\circ\pi_2)\le \delta
    		\, \implies \, W_1^{(\sigma)}(\pi_1, \pi_2)\le c_1\delta\log\frac{1}{\delta}.
    	\end{equation}
    \end{lemma}

    Applying Lemma~\ref{lmm:tv_w1} to~\eqref{eq:xxd1} we get  
    $$\sup_{\pi\in\mathcal{P}([0, a])}\bbE[W_1^{(\sigma)}(\pi, \hat{\pi})]\le n^{-1/4+o(1)}\,,$$
    which completes the proof of \prettyref{cor:got}.

    \begin{remark} [On the level of smoothing] We have obtained the bound for $\sigma$-smoothed distance between NPMLE and truth which is $n^{-c}$ for any $c<1/4$. This result required a constant fixed $\sigma$. However, it turns out that it is sufficient to set $\sigma = 1/\mathsf{Polylog}(n)$, while holding $a,\gamma$ fixed and letting $n\to\infty$.  Indeed, inspecting the proofs, 
     	the constant $c(a, \sigma, \gamma)$ in \prettyref{thm:poi_gauss} is $\exp(\frac{\sigma^2\gamma^2}{2})\cdot \mathsf{Poly}(\sigma, \frac{1}{\sigma})$.  On the other hand, 
        setting  $\frac{1}{\sigma^2}$ to grow with $(\log\frac{1}{\epsilon})^v$ where $0 < v < 1$, 
    	$t_{a, \sigma, \gamma}(\epsilon)$ becomes $\epsilon^{-v/2}$. Thus, the overall bound in RHS of the Theorem becomes $\epsilon^{(1-v)/2} \mathsf{Polylog}(\epsilon)$.
     Recalling that $\epsilon = \frac{1}{\sqrt{n}}\sqrt{\frac{\log n}{\log \log n}}$ we get the claimed $n^{-c}$ bound by taking $v$ sufficiently small.
    \end{remark}

    \section{Complex-analytic preliminaries}\label{sec:poi_gauss}

    The main proof technique for this work is complex analysis. Here, we remind that the $z$-transform $\mathcal{Z}(\pi)(z)$ (for priors $\pi$ with discrete support), 
    Laplace transform $\mathcal{L}(\pi)(s)$, and characteristic function $\Psi_{\pi}(t)$ of a distribution $\pi$ are defined as follows. 
    \begin{equation}
    	\mathcal{Z}(\pi)(z)\triangleq \sum_{n=0}^{\infty} \mathsf{PMF}(\pi)(n)z^n
    	\qquad\forall z\in\bbC
    \end{equation}
    \begin{equation}
    	\mathcal{L}(\pi)(s)\triangleq \bbE_{X\sim \pi}[\exp(sX)]
    	\qquad\forall s\in\bbC
    \end{equation}
    \begin{equation}
    	\Psi_{\pi}(t)\triangleq \bbE_{X\sim\pi}[\exp(itX)] = \mathcal{L}(\pi)(it)
    	\qquad\forall t\in \bbR
    \end{equation}

    We now consider the following identities for all bounded priors $\pi\in\mathcal{P}([0, a])$: for Poisson mixtures and Gaussian mixtures we have 
    \begin{equation}\label{eq:z_poi}
    	\mathcal{Z}(\Poi_{\gamma}\circ \pi)(z) = \mathcal{L}(\pi)(\gamma(z - 1))\quad\forall z\in\bbR\,;
    \end{equation}
    \begin{equation}\label{eq:lap_gauss}
    	\mathcal{L}(\Gsn_{\sigma}\circ \pi)(s)
    	=\exp\left(\frac{s^2\sigma^2}{2}\right)\mathcal{L}(\pi)(s)\quad\forall s\in\bbR\,.
    \end{equation}

    In addition, the Plancherel's theorem~\cite[Theorem 2]{wiener_fourier_1988} implies the following: 
    \begin{IEEEeqnarray}{Cl}\label{eq:gauss_l2}
    	&L_2(\Gsn_{\sigma}\circ \pi_1, \Gsn_{\sigma}\circ \pi_2)^2
    	\nonumber\\
    	\triangleq& 
    	\int_{-\infty}^{\infty}
    	((\Gsn_{\sigma}\circ \pi_1)(t) - (\Gsn_{\sigma}\circ \pi_2)(t))^2 dt
    	\nonumber\\
    	=&\frac{1}{2\pi}\int_{-\infty}^{\infty}
    	\exp(-\sigma^2t^2) |\Psi_{\pi_1}(t) - \Psi_{\pi_2}(t)|^2 dt
    \end{IEEEeqnarray}
    using the fact that the Fourier transform of the function $f(t)\triangleq e^{\frac{-\sigma^2t^2}{2}}\Psi_{\pi}(t)$ is 
    $\hat{f}(u) = 2\pi(\Gsn_{\sigma}\circ\pi)(2\pi u)$. 
    
    We now describe a main idea that we will be using: the Hadamard's three-circle theorem ~\cite[Theorem 12.1]{simon_2011} that states the following. 
    Let $x_0\in\bbC$, $r_0 < r_1\in\bbR$.
    Consider a function $f$ that is analytic on the annulus $A_{r_0, r_1} \triangleq \{z: r_0 < |z - x_0| < r_1\}$ and continuous everywhere else. 
    Denote $M_{x_0}(r; f) \triangleq \sup_{|z - x_0| \le r} |f(z)|$. 
    Then 
    \begin{equation}\label{eq:hadamard}
    	\text{$\log M_{x_0}(r; f)$ is a convex function of $\log r$.}
    \end{equation}
    Finally, we will also frequently use the following tail bound of the Gaussian distribution \cite[Theorem 4.7]{wasserman2004}. 
    \begin{equation}\label{eq:gsn_tail}
    	\bbP(N(0, \sigma) > T)\le \sqrt{\frac{2}{\pi}}\frac{\sigma\exp(-T^2/(2\sigma^2))}{T}, \quad \forall T > 0\,.
    \end{equation}
    
    We will use \prettyref{eq:hadamard} to bound the difference in characteristic functions of the Gaussian mixtures. 
    Then the $L_2$ distance can be bounded via \prettyref{eq:gauss_l2} and finally the TV distance via 
    \prettyref{lmm:gauss_l2tv}.

    \section{Proof of \prettyref{thm:poi_gauss}}
    The following lemma shows that it suffices to bound the $L_2$ distance in establishing \prettyref{thm:poi_gauss}. 
    
    \begin{lemma}\label{lmm:gauss_l2tv}
        Let $\epsilon, a > 0$ be given, $\pi_1$ and $\pi_2\in \mathcal{P}([0,a])$ be such that 
        \begin{equation}
            L_2(\Gsn_{\sigma}\circ\pi_1, \Gsn_{\sigma}\circ \pi_2)\le \epsilon
        \end{equation}
        Then 
        \begin{equation}
            \TV(\Gsn_{\sigma}\circ\pi_1, \Gsn_{\sigma}\circ\pi_2)\lesssim \epsilon\cdot\sqrt[4]{\sigma^2\log \frac{1}{\epsilon} + a}
        \end{equation}
    \end{lemma}
    \ifmapx 
    The complete proof is established in \prettyref{app:proofs_appendix}. 
    \else 
    \fi
    The proof idea is to bound the quantity $\int_{-T}^T |\Gsn_{\sigma}\circ \pi_1 - \Gsn_{\sigma}\circ\pi_2|dt$ using Cauchy–Schwarz inequality, and bound this quantity outside the said interval using \prettyref{eq:gsn_tail} and compactness of support. 

    Here, we consider the following lemma on transforming bounds on Laplace transform into the characteristic function, 
    relying only on the total variation of the Poisson mixtures and the support bound of the priors. 
    \begin{lemma}\label{lmm:epsilon_bound}
        Let $\pi_1, \pi_2\in \mathcal{P}([0,a])$ be such that
        \begin{equation}
            \sup_{|s + \gamma| \le \gamma} |\mathcal{L}(\pi_1)(s) - \mathcal{L}(\pi_2)(s)|\le 2\epsilon\,.
        \end{equation}
        Denote $R_\epsilon > 1$ a solution of 
        \begin{equation}\label{eq:r_choice}
            \log(1/\epsilon) = a(R_{\epsilon}(\log R_{\epsilon} - \log \gamma - 1) + \gamma)\,.
        \end{equation}
        Then for all $t \in \mathbb{R}$ we have
        \begin{equation}
            |\Psi_{\pi_1}(t) - \Psi_{\pi_2}(t)|\le 2\min\left(1, \epsilon\cdot\exp({a\over2} R_{\epsilon} \log(1
	    + \frac{t^2}{\gamma^2}))\right)\,.
        \end{equation}
    \end{lemma}

    \begin{IEEEproof}[Proof of \prettyref{lmm:epsilon_bound}]
    	Denote $f(s) = \frac{\mathcal{L}(\pi_1)(s)}{2} - \frac{\mathcal{L}(\pi_2)(s)}{2}$. 
    	For all $r > 0$, we consider 
    	$M(r) = \sup_{|s + \gamma|\le r} |f(s)|$ 
    	as per \prettyref{fig:circles}. 
    	Then we have the following estimates for $M$:
    	\begin{equation}
    		M(\gamma)\le \epsilon, 
    		\qquad 
    		\forall r > \gamma: M(r)\le \exp(a(r - \gamma))
    	\end{equation}
        where the second one is due to the fact that $\pi_1, \pi_2\in\mathcal{P}([0,a])$ 
        and $\sup_{|s+\gamma|\le r, x\in [0,a]} |\exp(sx)| =\exp(a(r-\gamma))$. 
    	Consider, now, the function $g(u) = \log(M(\gamma e^u))$, 
    	then we have $g(0)\le -\log(1/\epsilon)$ and for all $u > 0$, $g(u) \le a\gamma (e^u - 1)$.

        \pgfplotsset{compat=newest}
        
        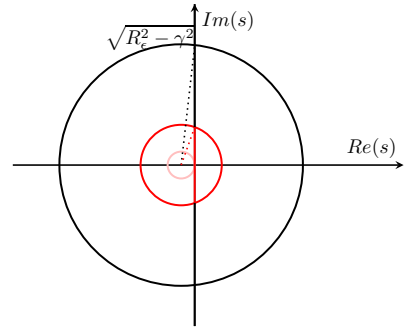
\begin{figure}[htbp]
        	\centering
        	\begin{tikzpicture}[scale=0.75]
        		\begin{axis}[
        			axis lines = left,
        			axis line style = thick,
        			xmin=-8, xmax=10, ymin=-12, ymax=12,
        			axis equal,
        			axis lines = middle,
        			ticks=none,
        			xlabel = $Re(s)$,
        			ylabel = {$Im(s)$},
        			xtick={},
        			xticklabels={},
        			ytick={13},
        			yticklabels = {$\sqrt{R_{\epsilon}^2 - \gamma^2}$},
        			]
        			\draw [line width=1pt, color=pink] (-1,0) circle [radius=1];
        			\draw [line width=1pt, color=red] (-1,0) circle [radius=3];
        			\draw [line width=1pt] (-1,0) circle [radius=9];
        			\draw [line width=1pt, color=red](axis cs: 0,-2.82) -- (axis cs:0,2.82);
        			\draw [line width=1pt, color=black](axis cs: 0,-12) -- (axis cs:0, -2.82);
        			\draw [line width=1pt, color=black](axis cs: 0, 2.82) -- (axis cs:0,12);
        			\draw [thick, dotted] (axis cs: -1,0) -- (axis cs: 0, 8.9);
        			\draw [thick, dotted, color=red] (axis cs: -1,0) -- (axis cs: 0, 2.82);
        			\draw[color=black] (0, 1)  node {};
        			\draw[color=black] (0, -4)  node {};
        			\draw[color=black] (-3.3, 9.5)  node {$\sqrt{R_{\epsilon}^2 - \gamma^2}$};
        		\end{axis}
        	\end{tikzpicture}
        	\caption{Bounding $|\Psi_{\pi_1} - \Psi_{\pi_2}|$ on red line using $M(r)$ on pink and black circles. }
        	\label{fig:circles}
        \end{figure}

        \begin{figure}[htbp]
        	\begin{center}
        		\begin{tikzpicture}[scale=0.75]
        			\begin{axis}[
        				xtick = {2},
        				xticklabels = {$u_{\epsilon}$},
        				xmin = -0.1, xmax = 3.1, ymin=-20, ymax=14,
        				ytick={-16.78},
        				yticklabels = {$\log(\epsilon)$},
        				axis lines = middle,
        				x label style={at={(axis description cs:1,0.7)},anchor=north},
        				y label style={at={(axis description cs:0,1.1)},anchor=north},
        				xlabel=$u$,
        				ylabel=$g(u)$,
        				legend pos=south east,
        				]
        				\addplot[
        				color=blue,
        				domain = 0:2.5,
        				] {2 * (exp(x) - 1)};
        				\addplot[
        				domain = 0:2.5,
        				] {2 * (exp(2) - 1) + (2 * exp(2)) * (x - 2)};
        				\node[circle,fill,inner sep=1.5pt] at (axis cs:0,-16.78) {};
        				\node[circle,fill,inner sep=1.5pt] at (axis cs:2, 12.778) {};
        				\legend {$a\gamma(e^u-1)$, tangent line}
        			\end{axis}
        		\end{tikzpicture}
        	\end{center}
        	\caption{Bound on $g(u)$ via Hadamard's 3-circle theorem.}
        	\label{fig:tangentplot}
        \end{figure}
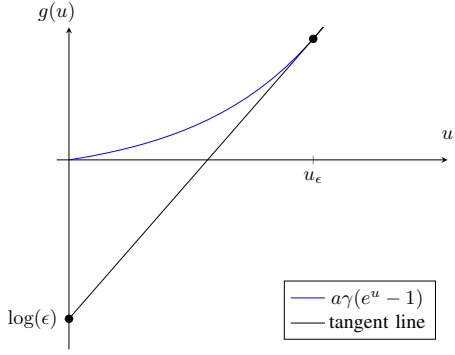
    	Given that both $\pi_1$ and $\pi_2$ are in $\mathcal{P}([0,a])$, 
    	$f$ is analytic on $\bbC$. Therefore, $g$ is convex by \prettyref{eq:hadamard}. 
    	Consider $R_{\epsilon}$ as given in \prettyref{eq:r_choice}. Let $u_{\epsilon} = \log(R_{\epsilon}) - \log(\gamma)$, 
    	then $\log(1/\epsilon) = a\gamma((u_{\epsilon} - 1)\exp(u_{\epsilon}) + 1)$. 
    	The motivation of this choice of $R_{\epsilon}$ and $u_\epsilon$ is given in
	\prettyref{fig:tangentplot}: for any choice of $u_\epsilon$ we would get an upper bound on
	$g$ given by the line joining the endpoints; the tangent line has the smallest slope, and therefore the best bound.
	
        For each $u\in [0, u_{\epsilon}]$, the convexity of $g$ entails 
    	\begin{IEEEeqnarray}{rCl}
    		g(u)&\le & g(0)(1 - \frac{u}{u_{\epsilon}}) + g(u_{\epsilon})\cdot \frac{u}{u_{\epsilon}}
    		\nonumber\\
    		&\le & -\log(1/\epsilon) + \frac{u}{u_{\epsilon}}(a\gamma(e^{u_{\epsilon}} - 1) + a\gamma((u_{\epsilon} - 1)e^{u_{\epsilon}} + 1))
    		\nonumber\\
    		&= & -\log(1/\epsilon) + a\gamma u\exp(u_{\epsilon})\,.
    	\end{IEEEeqnarray}

    	Now, $\Psi_{\pi_1}(t) - \Psi_{\pi_2}(t) = 2f(it)$. 
    	Since $|\Psi_{\pi}(t)|\le 1$ for all $\pi$, $|f(it)|\le 1$. 
    	On the other hand, $|it + \gamma| = \sqrt{\gamma^2+t^2}$. 
    	Therefore, for all $|t|\le \sqrt{R_{\epsilon}^2 - \gamma^2}$, we have 
    	\begin{IEEEeqnarray}{rCl}
    		|f(it)|
    		&\le & M(\sqrt{t^2+\gamma^2}) 
    		\nonumber\\
    		&= & \exp\left(g(\frac 12\log(1 + \frac{t^2}{\gamma^2}))\right)
    		\nonumber\\
    		&\le &\exp\left(-\log(1/\epsilon) + \frac{a\gamma\exp(u_{\epsilon})\log(1+\frac{t^2}{\gamma^2})}{2}\right)
    		\nonumber\\
    		&=&\epsilon\cdot \exp\left(\frac{aR_{\epsilon}\log(1+\frac{t^2}{\gamma^2})}{2}\right)\,.
    	\end{IEEEeqnarray}
	On the other hand,  for $|t| > \sqrt{R_\epsilon^2 - \gamma^2}$ we have $\epsilon\cdot
	\exp(\frac{aR_{\epsilon}\log(1+\frac{t^2}{\gamma^2})}{2}) \ge \exp(a(R_\epsilon -\gamma)) > 1$, implying that the bound is
	trivially true then. Thus, $f(it) \le \min(1, \epsilon\cdot \exp(aR_{\epsilon}\log(1+t^2) / 2))$ for
	all $t$.
    \end{IEEEproof}

    \begin{IEEEproof}[Proof of \prettyref{thm:poi_gauss}]
    	We first establish the following bound via \prettyref{eq:z_poi}. 
    	\begin{IEEEeqnarray}{Cl}
    		&\sup_{s: |s + \gamma|\le \gamma} |\mathcal{L}(\pi_1)(s) - \mathcal{L}(\pi_2)(s)|
    		\nonumber\\
    		=&\sup_{z: |z|\le 1} |\mathcal{Z}(\Poi_{\gamma}\circ\pi_1)(z) - \mathcal{Z}(\Poi_{\gamma}\circ\pi_2)(z)|
    		\nonumber\\
    		\le& \sum_{n=0}^{\infty} |(\Poi_{\gamma}\circ\pi_1)(n) - (\Poi_{\gamma}\circ\pi_2)(n)|
    		\nonumber\\
    		=& 2\mathsf{TV}(\Poi_{\gamma}\circ\pi_1, \Poi_{\gamma}\circ\pi_2) 
    		\le 2\epsilon
    	\end{IEEEeqnarray}
    
    	Motivated by \prettyref{eq:gauss_l2}, 
    	we consider $R_{\epsilon}$ as per \prettyref{lmm:epsilon_bound} and 
    	denote $E(s) \triangleq -\sigma^2s + aR_{\epsilon}\log(1+\frac{s}{\gamma^2})$ for all $s>-\gamma^2$. Then $E$ is concave and attains
	its global maximum at $s=\frac{aR_{\epsilon}}{\sigma^2} - \gamma^2$, thus for all $t\in \mathbb{R}$ we have
	\begin{IEEEeqnarray}{rCl}\label{eq:Ebnd}
		E(t^2) &\le & E_{\max} 
		\nonumber\\
		&:= &aR_{\epsilon}(\log (aR_{\epsilon}) - \log(\sigma^2\gamma^2) - 1) + \sigma^2 \gamma^2
		\nonumber\\
		&=& \log(1/\epsilon) - a\gamma +
	aR_{\epsilon}\log(\frac{a}{\sigma^2\gamma}) + \sigma^2\gamma^2 \,.
    \end{IEEEeqnarray}

	This means we may now bound the squared $L_2$ distance as follows:
    	\begin{IEEEeqnarray}{Cl}
    		&L_2(\Gsn_{\sigma}\circ\pi_1, \Gsn_{\sigma}\circ\pi_2)^2
    		\nonumber\\
    		\stepa{=} & \frac{1}{2\pi}\int_{-\infty}^{\infty} \exp(-\sigma^2t^2)|\Psi_{\pi_1}(t) - \Psi_{\pi_2}(t)|^2dt
    		\nonumber\\
    		\stepb{\lesssim} & \int_{-\infty}^{\infty} \exp(-\sigma^2t^2)\cdot\min(1, \epsilon\cdot\exp(\frac{aR_{\epsilon}\log(1+\frac{t^2}{\gamma^2})}{2}))^2dt
    		\nonumber\\
    		\stepc{\le} & \left(\int_{|t|\le R_\epsilon}\epsilon^2\exp(E(t^2))dt
    		+\int_{|t|>R_\epsilon}\exp(-\sigma^2t^2)dt
    		\right)
    		\nonumber\\
    		\stepd{\lesssim} & R_\epsilon \epsilon^2\exp(E_{\max}) + \exp(-\sigma^2R_\epsilon^2)
		\label{eq:Ebnd2}
    	\end{IEEEeqnarray}
	where in (a) we used Plancherel~\prettyref{eq:gauss_l2}, in (b) we
	applied~\prettyref{lmm:epsilon_bound}, in (c) we split the integral into two parts and
	applied respective bounds from previous line, in (d) we used~\prettyref{eq:Ebnd} and \prettyref{eq:gsn_tail}.

	To proceed, we notice that the function $f(r) = r\log r$ has $f(\frac{y}{\log y}) = y(1 - \frac{\log \log y}{\log y})$ for all $y > e$, 
	so as $y\to \infty$ the solution to $f(r) = y$ has $r = (1+ o(1))\frac{y}{\log y}$. 
	This, together with \prettyref{eq:r_choice}, implies that $R_{\epsilon} = (1+o(1))
	\frac 1a\frac{\log(1/\epsilon)}{\log\log(1/\epsilon)}$ as $\epsilon\to 0$.  Then, the second term in~\eqref{eq:Ebnd2} is
	$o(\epsilon) = o_{a, \sigma, \gamma}(\epsilon)$ and can be neglected, whereas for the first term we can see 
	from~\prettyref{eq:Ebnd} that 
	$$ \exp{E_{\max}} = {1\over \epsilon} s_{a, \sigma, \gamma}(\epsilon)\,, $$
	$$s_{a, \sigma, \gamma}(\epsilon) := \exp\left\{\sigma^2\gamma^2 - a\gamma + \left(\log \frac{a}{\sigma^2\gamma} + o(1)\right) { \log {1\over \epsilon}
	\over \log \log {1\over \epsilon}} \right\}\,.$$

	Collecting terms, thus, we have shown that as $\epsilon \to 0$ we have
	$$ 
    		L_2(\Gsn_{\sigma}\circ\pi_1, \Gsn_{\sigma}\circ\pi_2)^2  \lesssim \epsilon R_\epsilon s_{a, \sigma, \gamma}(\epsilon) \,.
	$$

    	Finally, taking the square root and invoking \prettyref{lmm:gauss_l2tv} we obtain the
	statement of the theorem. 
    \end{IEEEproof}
    
    \section{Proof of \prettyref{thm:gauss_poi}}
    For the comparison in the other direction, we need the following bound on the magnitude of the difference of Laplace transform. 
    \begin{lemma}\label{lmm:laplace_bound}
    	Consider the same setting as before, where 
    	$\pi_1, \pi_2\in\mathcal{P}([0,a])$. 
    	Given $\epsilon > 0$ such that 
    	\begin{equation}
    		\TV(\Gsn_{\sigma}\circ\pi_1, \Gsn_{\sigma}\circ\pi_2)\le \epsilon\,.
    	\end{equation}
    	Then the Laplace transform satisfies the following: 
    	
    	\begin{IEEEeqnarray}{Cl}
    		&|\mathcal{L}(\pi_1)(s) - \mathcal{L}(\pi_2)(s)|
    		\nonumber\\
    		\lesssim &\epsilon\exp\left(-\frac{\sigma^2Re(s^2)}{2} + E_{a, \sigma}(\epsilon, s)\right)\,.
    	\end{IEEEeqnarray}
        \begin{equation}\label{eq: ea_sigma}
        	E_{a, \sigma}(\epsilon, s) := \sigma^2Re(s)^2 + a\cdot |Re(s)| + |Re(s)|\sqrt{2\sigma^2\log \frac{1}{\epsilon}}\,.
        \end{equation}
    \end{lemma}

    \begin{IEEEproof}[Proof of \prettyref{lmm:laplace_bound}]
    	We will show that 
    	\begin{IEEEeqnarray}{cl}\label{eq:laplace_bound}
    		&|\mathcal{L}(\Gsn_{\sigma}\circ\pi_1 - \Gsn_{\sigma}\circ\pi_2)(s)|
    		\le\epsilon\exp\left(E_{a, \sigma}(\epsilon, s)\right)
    	\end{IEEEeqnarray}
        with $E_{a, \sigma}(\epsilon, s)$ as per \prettyref{eq: ea_sigma}, 
        and then the conclusion follows from \prettyref{eq:lap_gauss}. 
    	
    	Indeed, we first consider the following: 
    	\begin{IEEEeqnarray}{Cl}
    		&|\mathcal{L}(\Gsn_{\sigma}\circ\pi_1 - \Gsn_{\sigma}\circ\pi_2)(s)|
    		\nonumber\\
    		\le& \int_{-\infty}^{\infty} 
    		|\exp(st)\cdot (\Gsn_{\sigma}\circ\pi_1(t) - \Gsn_{\sigma}\circ\pi_2(t))|dt
    		\nonumber\\
    		=&
    		\int_{-\infty}^{\infty}
    		\exp(Re(st)) |\Gsn_{\sigma}\circ\pi_1(t) - \Gsn_{\sigma}\circ\pi_2(t)| dt.\,
    	\end{IEEEeqnarray}
    	Consider $T > \sigma^2|Re(s)|+ a$, we now split this into three parts:
    	\[
    	\int_{-\infty}^{-T}
    	\qquad 
    	\int_{-T}^T
    	\qquad 
    	\int_{T}^{\infty}
    	\]
    	First, the term in the middle: 
    	\begin{IEEEeqnarray}{Cl}
    		&\int_{-T}^T
    		\exp(Re(st)) |\Gsn_{\sigma}\circ\pi_1(t) - \Gsn_{\sigma}\circ\pi_2(t)| dt
    		\nonumber\\
    		\le &
    		\sup_{|t|\le T}\exp(Re(st))
    		\int_{-T}^T
    		|\Gsn_{\sigma}\circ\pi_1(t) - \Gsn_{\sigma}\circ\pi_2(t)| dt
    		\nonumber\\
    		\le & \exp(T\cdot |Re(s)|)\epsilon \,.
    	\end{IEEEeqnarray}
    	Next, for each $\pi\in\mathcal{P}([0, a])$, 
    	$\Gsn_{\sigma}\circ\pi(t)$ is nonnegative for all $t$, 
    	while also bounded above by 
    	$\frac{1}{\sqrt{2\pi}\sigma}\exp(-\frac{-t^2}{2\sigma^2})$ for $t \le 0$, and $\frac{1}{\sqrt{2\pi}\sigma}\exp(-\frac{(t - a)^2}{2\sigma^2})$ for $t \ge a$. 
    	Therefore, denoting:
    	\begin{equation}
    		M_1(T) \triangleq T + \sigma^2Re(s),
    		\qquad 
    		M_2 (T) \triangleq T - a - \sigma^2Re(s)\,,
    	\end{equation}
    	the left tail can be computed as 
    	\begin{IEEEeqnarray*}{Cl}
    		&\int_{-\infty}^{-T}
    		\exp(Re(st)) |\Gsn_{\sigma}\circ\pi_1(t) - \Gsn_{\sigma}\circ\pi_2(t)| dt
    		\nonumber \\
    		\le & \frac{1}{\sqrt{2\pi}\sigma}
    		\int_{-\infty}^{-T}
    		\exp(t Re(s)) \cdot \exp(-\frac{t^2}{2\sigma^2})dt
    		\nonumber \\
    		\stepa{=} & \frac{1}{\sqrt{2\pi}\sigma}
    		\int_{-\infty}^{-T}
    		\exp(-\frac{(t- \sigma^2Re(s))^2}{2\sigma^2} + \frac{\sigma^2Re(s)^2}{2})dt
    		\nonumber \\
    		\stepb{\le} & \sqrt{\frac{2}{\pi}}\frac{\sigma}{M_1 (T)}\exp(\frac{\sigma^2Re(s)^2}{2})\cdot \exp(-\frac{M_1 (T)^2}{2\sigma^2})
    	\end{IEEEeqnarray*}
        where (a) is completing the square and (b) follows from \prettyref{eq:gsn_tail}. 
    
    	We also have the right tail computed similarly as 
    	\begin{IEEEeqnarray*}{cl}
    		&\int_{T}^{\infty}
    		\exp(Re(st)) |\Gsn_{\sigma}\circ\pi_1(t) -\Gsn_{\sigma}\circ\pi_2(t)| dt
    		\nonumber \\
    		\le &\frac{1}{\sqrt{2\pi}\sigma}
    		\int_{T}^{\infty}
    		\exp(t Re(s)) \cdot \exp(-\frac{(t-a)^2}{2\sigma^2})dt
    		\nonumber \\
    		\le &\sqrt{\frac{2}{\pi}}\frac{\sigma}{M_2(T)}\exp(\frac{\sigma^2Re(s)^2}{2} + a \cdot Re(s))
    		\cdot \exp(-\frac{M_2(T)^2}{2\sigma^2})\,.
    	\end{IEEEeqnarray*}

    	Denote, now, $M_3(T)\triangleq T - a - \sigma^2|Re(s)|$. 
    	Then $\min\{M_1(T), M_2(T)\}\ge M_3(T) \ge 0$. 
    	Therefore, collecting terms above, 
    	\begin{IEEEeqnarray}{cl}
    		&|\mathcal{L}(\Gsn_{\sigma}\circ\pi_1- \Gsn_{\sigma}\circ\pi_2)(s)|
    		\nonumber\\
    		 \lesssim &\epsilon\exp(T\cdot|Re(s)|) \label{eq:aa1}\\
    		{} + & \frac{\sigma}{M_3(T)}\exp(-\frac{M_3(T)^2}{2\sigma^2}
    		+\frac{\sigma^2Re(s)^2}{2} + a\cdot |Re(s)|). \label{eq:aa2}
    	\end{IEEEeqnarray}
    	
    	Next, we choose $T = \sigma^2|Re(s)| + a + \sqrt{2\sigma^2\log \frac{1}{\epsilon}}$. Thus, the first
	term~\eqref{eq:aa1} evaluates to
    	\[
    	\epsilon\exp(\sigma^2Re(s)^2 + a\cdot|Re(s)| + |Re(s)|\sqrt{2\sigma^2\log \frac{1}{\epsilon}})\,.
    	\]
    	With this choice of $T$, we have  $M_3(T) = \sqrt{2\sigma^2\log\frac{1}{\epsilon}}$.
	Then, the second term~\eqref{eq:aa2} is bounded as
    	\[
    	\frac{\sigma}{\sqrt{2\sigma^2\log(\frac{1}{\epsilon})}}
    	\exp(\frac{\sigma^2Re(s)^2}{2} + a\cdot |Re(s)|)\epsilon\,.
    	\]
    	Therefore collecting the two terms together, and taking the maximum of the exponents, gives us 
    	\prettyref{eq:laplace_bound}. 
    	
    \end{IEEEproof}
    
    \begin{IEEEproof}[Proof of \prettyref{thm:gauss_poi}] As in 
    	\cite[(33)]{polyanskiy_sample_2017} we use the standard fact that for any real $r > 1$, 
    	a function $f(z)\triangleq \sum_{n=0}^{\infty} a_nz^n$ satisfies 
    	\begin{equation}
    		\sum_{n=0}^{\infty} |a_n|\le \frac{r}{r - 1}\sup_{|z|\le r} |f(z)|
    	\end{equation}
        (as a consequence of Cauchy's integral formula).

    	Now if $a_n = (\Poi\circ\pi_1)(n) - (\Poi\circ\pi_2)(n)$, then using \prettyref{eq:z_poi}, 
    	$f(z) = \mathcal{L}(\pi_1)(\gamma(z - 1)) - \mathcal{L}(\pi_2)(\gamma(z - 1))$. 
    	Thus setting $r = 2$, we have, by \prettyref{lmm:laplace_bound}, 
    	\begin{IEEEeqnarray*}{Cl}
    		&2\TV(\Poi\circ\pi_1, \Poi\circ\pi_2)
    		\nonumber\\
    		\le& 2\sup_{|z|\le r} |f(z)|
    		\nonumber\\
    		=&2\sup_{|z|\le r}|\mathcal{L}(\pi_1)(\gamma(z - 1)) - \mathcal{L}(\pi_2)(\gamma(z - 1))|
    		\nonumber\\
    		\lesssim &\sup_{|s+\gamma|\le 2\gamma} \epsilon
    		\exp\left(-\frac{\sigma^2Re(s^2)}{2} + E_{a, \sigma}(\epsilon, s)\right)
    		\nonumber\\
    		\le &\epsilon\exp\left(3\gamma\sigma\cdot\sqrt{2\log \frac{1}{\epsilon}}
    		+\frac{9\gamma^2\sigma^2}{2} + 9\gamma^2\sigma^2 + 3\gamma a\right)
    	\end{IEEEeqnarray*}
    	where we used 
    	$|Re(s)|\le |s|\le 3\gamma$ 
    	and $|Re(s^2)|\le |s^2|\le 9\gamma^2$
    	for all $s$ with $|s+\gamma|\le 2\gamma$.  
    	
    \end{IEEEproof}

    \section*{Acknowledgments}
    This material is based upon work supported by the National Science Foundation under Grant No CCF-2131115. Anzo Teh was supported by a fellowship from the Eric and Wendy Schmidt Center at the Broad Institute.

    \bibliographystyle{IEEEtran}
    \bibliography{references}

\begin{thebibliography}{10}
\providecommand{\url}[1]{#1}
\csname url@samestyle\endcsname
\providecommand{\newblock}{\relax}
\providecommand{\bibinfo}[2]{#2}
\providecommand{\BIBentrySTDinterwordspacing}{\spaceskip=0pt\relax}
\providecommand{\BIBentryALTinterwordstretchfactor}{4}
\providecommand{\BIBentryALTinterwordspacing}{\spaceskip=\fontdimen2\font plus
\BIBentryALTinterwordstretchfactor\fontdimen3\font minus
  \fontdimen4\font\relax}
\providecommand{\BIBforeignlanguage}[2]{{%
\expandafter\ifx\csname l@#1\endcsname\relax
\typeout{** WARNING: IEEEtran.bst: No hyphenation pattern has been}%
\typeout{** loaded for the language `#1'. Using the pattern for}%
\typeout{** the default language instead.}%
\else
\language=\csname l@#1\endcsname
\fi
#2}}
\providecommand{\BIBdecl}{\relax}
\BIBdecl

\bibitem{bash2013limits}
B.~A. Bash, D.~Goeckel, and D.~Towsley, ``Limits of reliable communication with
  low probability of detection on awgn channels,'' \emph{IEEE journal on
  selected areas in communications}, vol.~31, no.~9, pp. 1921--1930, 2013.

\bibitem{goldfeld_convergence_2020}
Z.~Goldfeld, K.~Greenewald, Y.~Polyanskiy, and J.~Weed, ``Convergence of
  {Smoothed} {Empirical} {Measures} with {Applications} to {Entropy}
  {Estimation},'' in \emph{IEEE Trans. Inf. Theory}, vol.~66, no.~7, Jul. 2020,
  pp. 4368--4391.

\bibitem{han_nonparametric_2021}
F.~Han, Z.~Miao, and Y.~Shen, ``Nonparametric mixture {MLEs} under
  {Gaussian}-smoothed optimal transport distance,'' \emph{arXiv preprint
  arXiv:2112.02421}, Dec. 2021.

\bibitem{miao_fisher-pitman_2021}
Z.~Miao, W.~Kong, R.~K. Vinayak, W.~Sun, and F.~Han, ``Fisher-{Pitman}
  permutation tests based on nonparametric {Poisson} mixtures with application
  to single cell genomics,'' \emph{arXiv preprint arXiv:2106.03022}, Jun. 2021.

\bibitem{jana_optimal_2022}
S.~Jana, Y.~Polyanskiy, and Y.~Wu, ``Optimal empirical {Bayes} estimation for
  the {Poisson} model via minimum-distance methods,'' \emph{arXiv preprint
  arXiv:2209.01328}, Sep. 2022.

\bibitem{MS13}
A.~Moitra and M.~Saks, ``A polynomial time algorithm for lossy population
  recovery,'' in \emph{Foundations of Computer Science (FOCS), 2013 IEEE 54th
  Annual Symposium on}.\hskip 1em plus 0.5em minus 0.4em\relax IEEE, 2013, pp.
  110--116.

\bibitem{polyanskiy_sample_2017}
\BIBentryALTinterwordspacing
Y.~Polyanskiy, A.~T. Suresh, and Y.~Wu, ``Sample complexity of population
  recovery,'' in \emph{Proceedings of the 2017 Conference on Learning Theory},
  ser. Proceedings of Machine Learning Research, S.~Kale and O.~Shamir, Eds.,
  vol.~65.\hskip 1em plus 0.5em minus 0.4em\relax PMLR, 07--10 Jul 2017, pp.
  1589--1618. [Online]. Available:
  \url{https://proceedings.mlr.press/v65/polyanskiy17a.html}
\BIBentrySTDinterwordspacing

\bibitem{dual2-2019}
Y.~Polyanskiy and Y.~Wu, ``Dualizing {L}e {C}am's method for functional
  estimation, with applications to estimating the unseens,'' \emph{arXiv
  preprint arXiv:1902.05616}, Feb. 2019.

\bibitem{jana2020extrapolating}
S.~Jana, Y.~Polyanskiy, and Y.~Wu, ``Extrapolating the profile of a finite
  population,'' in \emph{Conference on Learning Theory}.\hskip 1em plus 0.5em
  minus 0.4em\relax PMLR, 2020, pp. 2011--2033.

\bibitem{polyanskiy_information_2022}
\BIBentryALTinterwordspacing
Y.~Polyanskiy and Y.~Wu, \emph{Information {Theory}: {From} {Coding} to
  {Learning}}.\hskip 1em plus 0.5em minus 0.4em\relax Cambridge University
  Press, 2022+. [Online]. Available:
  \url{https://people.lids.mit.edu/yp/homepage/data/itbook-export.pdf}
\BIBentrySTDinterwordspacing

\bibitem{adell2006exact}
J.~A. Adell and P.~Jodrá, ``\BIBforeignlanguage{en}{Exact {Kolmogorov} and
  total variation distances between some familiar discrete distributions},''
  \emph{\BIBforeignlanguage{en}{Journal of Inequalities and Applications}},
  vol. 2006, no.~1, pp. 1--8, Dec. 2006, number: 1 Publisher: SpringerOpen.

\bibitem{eb2021}
Y.~Polyanskiy and Y.~Wu, ``Sharp regret bounds for empirical {B}ayes and
  compound decision problems,'' \emph{arXiv preprint arXiv:2109.03943}, Sep.
  2021.

\bibitem{lambert_asymptotic_1984}
D.~Lambert and L.~Tierney, ``Asymptotic {Properties} of {Maximum} {Likelihood}
  {Estimates} in the {Mixed} {Poisson} {Model},'' \emph{The Annals of
  Statistics}, vol.~12, no.~4, pp. 1388--1399, Dec. 1984, publisher: Institute
  of Mathematical Statistics.

\bibitem{optimal.transport.old.new}
C.~Villani, \emph{{Optimal Transport: Old and New}}.\hskip 1em plus 0.5em minus
  0.4em\relax Berlin: Springer Verlag, 2008.

\bibitem{wiener_fourier_1988}
N.~Wiener, \emph{\BIBforeignlanguage{eng}{The Fourier integral and certain of
  its applications}}, ser. Cambridge mathematical library.\hskip 1em plus 0.5em
  minus 0.4em\relax Cambridge: Cambridge University Press, 1988.

\bibitem{simon_2011}
B.~Simon, \emph{Convexity: An Analytic Viewpoint}, ser. Cambridge Tracts in
  Mathematics.\hskip 1em plus 0.5em minus 0.4em\relax Cambridge University
  Press, 2011.

\bibitem{wasserman2004}
\BIBentryALTinterwordspacing
L.~Wasserman, \emph{Inequalities}.\hskip 1em plus 0.5em minus 0.4em\relax New
  York, NY: Springer New York, 2004, pp. 63--69. [Online]. Available:
  \url{https://doi.org/10.1007/978-0-387-21736-9_4}
\BIBentrySTDinterwordspacing

\end{thebibliography}

    \ifmapx
    \appendices

    \section{Proofs of Auxillary Lemmas}\label{app:proofs_appendix}
    \begin{IEEEproof}[Proof of Corollary~\ref{cor:got}]
    	We first consider the following steps: there is a constant $c_1= c_1(a)$ such that 
    	\begin{IEEEeqnarray}{Cl}\label{eq:tv_npmle}
    		&\sup_{\pi\in\mathcal{P}([0,a])} \bbE[\TV(\Poi\circ\pi, \Poi\circ\hat{\pi})]
    		\nonumber\\
    		\stepa{\le} &\sup_{\pi\in\mathcal{P}([0,a])} \bbE[H(\Poi\circ\pi, \Poi\circ\hat{\pi})]
    		\nonumber\\
    		\stepb{\le}&\sup_{\pi\in\mathcal{P}([0,a])} \sqrt{\bbE[H^2(\Poi\circ\pi, \Poi\circ\hat{\pi})]}
    		\nonumber\\
    		\stepc{\lesssim}& \sup_{\pi\in\mathcal{P}([0,a])} c_1(a)\frac{1}{\sqrt{n}}\cdot\sqrt{\frac{\log n}{\log \log n}}
    	\end{IEEEeqnarray}
    	where (a) is due to $\TV(P, Q)\le H(P, Q)$ ~\cite[(7.20)]{polyanskiy_information_2022}, 
    	(b) is $ (\bbE[H(P, Q)])^2\le \bbE[H^2(P, Q)]$ by Cauchy-Schawrz inequality, 
    	and (c) is by~\eqref{eq:jana}.

    	Combining \prettyref{thm:poi_gauss} and \prettyref{lmm:tv_w1}, we see that there is a constant $c_2= c_2(a, \sigma)$ such that for all $\pi, \hat{\pi}\in \mathcal{P}([0, a])$ and $X > 0$, 
    	\begin{equation}\label{eq:main_and_lemma}
    		\TV(\Poi\circ\pi, \Poi\circ\hat{\pi})\le X
    		\implies
    		W_1^{(\sigma)}(\pi, \hat{\pi}) \le c_2\sqrt{X} u_{a, \sigma}(X)
    	\end{equation}
    	where for all $x$ with $0 < x < \frac {1}{2e}$ we define 
    	$u_{a, \sigma}(x)\triangleq t_{a, \sigma}(x) \log\left(\frac{1}{x}\right)$, 
    	$t_{a, \sigma}(x)$ as per \prettyref{thm:poi_gauss}. 
    	
    	Now we have two cases: 
    	\begin{itemize}
    		\item If $a < \sigma^2$, then $\lim u_{a, \sigma}(x)\to 0$ as $x \to 0$, 
    		(the polylog factor of $\frac{1}{x}$ is offset by the 
    		factor $\exp((\frac{\log a - \log \sigma^2}{2} + o(1))\frac{\log \frac 1x}{\log \log \frac 1x})$)
    		so $u_{a, \sigma}(x)$ is bounded in $(0, \frac 2e)$ by some factor $C=C(a, \sigma)$. 
    		
    		\item If $a\ge \sigma^2$, then there is $N=N(a, \sigma^2)$ such that 
    		$u_{a,\sigma}(\frac{1}{\sqrt{n}})$ is increasing in $n$ but $\frac{1}{\sqrt[4]{n}}u_{a,\sigma}(\frac{1}{\sqrt{n}})$ is decreasing in $n$ for $n \ge N$. 
    		This means, 
    		when $X\ge \frac{1}{\sqrt{n}}$, $u_{a, \sigma}(X)\le u_a(\frac {1}{\sqrt{n}})$;
    		when $X\le \frac{1}{\sqrt{n}}$, $\sqrt{X}u_{a, \sigma}(X)\le \frac{1}{\sqrt[4]{n}}u_a(\frac {1}{\sqrt{n}})$.
    	\end{itemize}
    	
    	The first case gives us 
    	\begin{IEEEeqnarray*}{C}\label{eq:exp_x}
    		\sup_{\pi\in\mathcal{P}([0,a])}\bbE[W_1^{(\sigma)}(\pi, \hat{\pi})]
    		\le C\bbE[\sqrt{X}]
    		\nonumber\\
    		\stepa{\le} C\sqrt{\bbE[X]}
    		\stepb{\lesssim} Cc_2(\frac{\log n}{n\log \log n})^{1/4}
    	\end{IEEEeqnarray*}
    	where (a) follows from Cauchy-Schwarz inequality and 
    	(b) from \prettyref{eq:tv_npmle}.

    	For the second case, we have 
    	\begin{IEEEeqnarray}{rCl}
    		\sup_{\pi\in\mathcal{P}([0,a])}\bbE[W_1^{(\sigma)}(\pi, \hat{\pi})] 
    		&\stepa{\le} &\bbE[\sqrt{X} u_{a, \sigma}(X)]
    		\nonumber\\
    		&= &\bbE[\sqrt{X} u_{a, \sigma}(X)\indc{X \ge \frac{1}{\sqrt{n}}}]
    		\nonumber\\
    		&+&\bbE[\sqrt{X} u_{a, \sigma}(X)\indc{X < \frac{1}{\sqrt{n}}}]
    		\nonumber\\
    		&\le  &u_{a, \sigma}\left(\frac {1}{\sqrt{n}}\right)\bbE[\sqrt{X}]
    		+ \frac{1}{\sqrt[4]{n}} u_{a, \sigma}\left(\frac {1}{\sqrt{n}}\right)
    		\nonumber\\
    		&\stepb{\lesssim} &c_2\left(\frac{\log n}{n\log \log n}\right)^{1/4} u_{a, \sigma}\left(\frac {1}{\sqrt{n}}\right)
    	\end{IEEEeqnarray}
    	where the (a)  follows from \prettyref{eq:main_and_lemma}, 
    	and (b) from Cauchy-Schawrz and \prettyref{eq:tv_npmle}. 
    	Finally, 
    	\begin{IEEEeqnarray*}{Cl}
    		&u_{a, \sigma}(\frac{1}{\sqrt{n}})
    		\nonumber\\
    		=&\frac{\log(\sqrt{n})^{7/4}}{\sqrt{\log\log(\sqrt{n})}}\exp(({\log a - \log(\sigma^2) \over 2} + o(1)) { \log(\sqrt{n})\over \log \log(\sqrt{n})}) 
    	\end{IEEEeqnarray*}
    	which is $n^{o_{a, \sigma}(1)}$ as $n\to\infty$. Therefore $\bbE[W_1^{(\sigma)}(\pi, \hat{\pi})] \lesssim n^{-1/4+o_{a, \sigma}(1)}$. 
    \end{IEEEproof}
    
    \begin{IEEEproof}[Proof of \prettyref{lmm:tv_w1}]
    	We consider the following statement in \cite[Theorem 6.15]{optimal.transport.old.new} (using $p=1, p'=\infty$) : 
    	For any point $x_0$ we have 
    	\begin{equation}
    		W_1(\pi_1, \pi_2) \le \int_{x=-\infty}^{\infty} |x_0-x| |d\pi_1(x) - d\pi_2(x)|\,.
    	\end{equation}
    	
    	Choose $x_0 = \frac a2$, then for every $T > a$ we have 
    	
    	\begin{IEEEeqnarray}{Cl}\label{eq:w1_bound}
    		&W_1^{(\sigma)}(\pi_1, \pi_2)
    		\nonumber\\
    		\le &\int_{-\infty}^{\infty} |u - \frac{a}{2}| \!\cdot \!
    		|(\Gsn_{\sigma}\circ \pi_1)(u) - (\Gsn_{\sigma}\circ\pi_2)(u)|du
    		\nonumber\\
    		\le & \int_{|u - \frac a2|\le T - \frac a2} |u- \frac a2|\!\cdot \!
    		|(\Gsn_{\sigma}\circ \pi_1)(u) - (\Gsn_{\sigma}\circ\pi_2)(u)|du
    		\nonumber\\
    		+ &\int_{|u - \frac a2|> T - \frac a2} |u- \frac a2| \!\cdot \!
    		|(\Gsn_{\sigma}\circ \pi_1)(u) - (\Gsn_{\sigma}\circ\pi_2)(u)|du
    		\nonumber\\
    		\le &2\delta|T - \frac a2|^{2}
    		\nonumber\\
    		+&\int_{|u - \frac a2|> T - \frac a2} |u- \frac a2| \!\cdot \!
    		|(\Gsn_{\sigma}\circ \pi_1)(u) - (\Gsn_{\sigma}\circ\pi_2)(u)|du\,.
    		\nonumber\\
    	\end{IEEEeqnarray}
    	Because $\pi_1$ and $\pi_2$ are supported on $[0, a]$, 
    	for $\pi\in \{\pi_1,\pi_2\}$ and for all $u$ with $|u - \frac a2| > \frac a2$ we have 
    	\[
    	0\le \Gsn_{\sigma}\circ \pi(u)
    	\le \frac{1}{\sqrt{2\pi}\sigma} \exp(-\frac{(|u-a/2| -  a/2)^2}{2\sigma^2})\,.
    	\]
    	Now that the tail bound is symmetric on both sides, we have 
    	\begin{IEEEeqnarray}{Cl}
    		&\int_{|u - \frac a2|> T - \frac a2} |u - \frac a2|\!\cdot\!
    		|(\Gsn_{\sigma}\circ \pi_1)(u) - (\Gsn_{\sigma}\circ\pi_2)(u)|du
    		\nonumber\\
    		\le& \frac{2}{\sqrt{2\pi}\sigma} \int_{T}^{\infty} |u - \frac a2|\exp(-\frac{(|u-a/2| -  a/2)^2}{2\sigma^2})du
    		\nonumber\\
    		=& \frac{2}{\sqrt{2\pi}\sigma} \int_{T}^{\infty} (u - \frac a2)\exp(-\frac{(u - a)^2}{2\sigma^2})du
    		\nonumber\\
    		\stepa{=} & a\bbP[N(0, \sigma^2) > T - a] + \frac{2\sigma}{\sqrt{2\pi}}\exp(-\frac{(T - a)^2}{\sqrt{2\pi}})
    		\nonumber\\
    		\stepb{\le} &\left(\frac{\sqrt{2}a\sigma}{\sqrt{\pi}(T - a)} + \frac{2\sigma}{\sqrt{2\pi}}\right)
    		\exp(-\frac{(T - a)^2}{2\sigma^2})
    	\end{IEEEeqnarray}
    	where (a) is by the expansion of 
    	$(u - \frac a2)\exp(-\frac{(u - a)^2}{2\sigma^2})$
    	into $\frac a2\exp(-\frac{(u - a)^2}{2\sigma^2}) + (u - a)\exp(-\frac{(u - a)^2}{2\sigma^2})$, and 
    	(b) (first term) is due to ~\cite[Theorem 4.7]{wasserman2004}.

    	Finally, setting $T = \sqrt{2\sigma^2\log(1/\delta)} + a$, 
    	\prettyref{eq:w1_bound} is now bounded by
    	\begin{IEEEeqnarray*}{Cl}
    	&2\left(\sqrt{2\sigma^2\log(1/\delta)} + \frac{a}{2}\right)^2\delta 
    	\\
    	+ &\left(\frac{\sqrt{2}a\sigma}{\sqrt{\pi}\sqrt{2\sigma^2\log(1/\delta)} } + \frac{2\sigma}{\sqrt{2\pi}}\right)\delta
    	\\
    	\lesssim &
    	\delta\left(2\sigma^2\log(1/\delta) + a^2 + a + \sigma\right)\,.
    	\end{IEEEeqnarray*}
    	
    \end{IEEEproof}
    
    
    \begin{IEEEproof}[Proof of \prettyref{lmm:gauss_l2tv}]
        According to the definition of L2, we have 
        \begin{equation}
            \int_{-\infty}^{\infty}
            ((\Gsn_{\sigma}\circ\pi_1)(t) - (\Gsn_{\sigma}\circ\pi_2(t))^2dt
            \le \epsilon^2.
        \end{equation}
        Consider any $T > a$. 
        By Cauchy-Schwarz inequality we have 
        \begin{IEEEeqnarray}{Cl}\label{eq:l2w1_cs}
        &\left(\int_{-T}^T |(\Gsn_{\sigma}\circ\pi_1)(t) - (\Gsn_{\sigma}\circ\pi_2)(t)|dt\right)^2
        \nonumber\\
        \le &\left(\int_{-T}^T ((\Gsn_{\sigma}\circ\pi_1)(t) - (\Gsn_{\sigma}\circ\pi_2)(t))^2 dt\right)
        \left(\int_{-T}^T 1 dt\right)
        \nonumber\\
        \lesssim & 2T\cdot \epsilon^2\,.
        \end{IEEEeqnarray}
        In addition, 
        since both $\pi_1$ and $\pi_2\in \mathcal{P}([0, a])$, 
        we have 
        $\bbP(|\Gsn_{\sigma}\circ\pi| > T) \le 2\bbP(N(0, \sigma) > T - a)\lesssim \frac{2\sigma}{T - a}\exp(-\frac{(T - a)^2}{2\sigma^2})$ for each $\pi\in \{\pi_1, \pi_2\}$, 
        with the last inequality follows from 
        \prettyref{eq:gsn_tail}. 
        This means 
        \begin{IEEEeqnarray}{Cl}\label{eq:l2w1_tail}
        	&\int_{|t| > T} |(\Gsn_{\sigma}\circ\pi_1)(t) - (\Gsn_{\sigma}\circ\pi_2)(t)|dt
        	\nonumber\\
        	\le &\int_{|t| > T} |(\Gsn_{\sigma}\circ\pi_1)(t)| + |(\Gsn_{\sigma}\circ\pi_2)(t)|dt
        	\nonumber\\
        	\lesssim &\frac{2\sigma}{T - a}\exp(-\frac{(T - a)^2}{2\sigma^2})\,.
        \end{IEEEeqnarray}
        
        Thus collecting \prettyref{eq:l2w1_cs} and \prettyref{eq:l2w1_tail} we have 
        \begin{IEEEeqnarray*}{Cl}
        &2\TV((\Gsn_{\sigma}\circ\pi_1)(t), (\Gsn_{\sigma}\circ\pi_2)(t))
        \nonumber\\
        \lesssim &\sqrt{T}\cdot \epsilon + \frac{\sigma}{T-a}
        \exp(-\frac{(T - a)^2}{2\sigma^2})\,.
        \end{IEEEeqnarray*}
        Now choose $T =  \sqrt{2\sigma^2\log(\frac{1}{\epsilon})} + a$, we have 
        \begin{IEEEeqnarray*}{Cl}
            &\sqrt{T}\cdot \epsilon + \frac{1}{T - a}\exp(-\frac{(T - a)^2}{2})
            \nonumber\\
            = &\epsilon\sqrt{\sqrt{2\sigma^2\log(\frac{1}{\epsilon})} + a}
            + \frac{\epsilon}{\sqrt{2\log(\frac{1}{\epsilon})}}
            \nonumber\\
            \lesssim &\epsilon\cdot\sqrt[4]{\sigma^2\log(\frac{1}{\epsilon}) + a}
        \end{IEEEeqnarray*}
        where the last inequality we used 
        $\sqrt{2\sigma^2\log(\frac{1}{\epsilon})} + a\le \sqrt{2(2\sigma^2\log(\frac{1}{\epsilon}) + a^2)}$. 
    \end{IEEEproof}

   \fi
   

\end{document}